\documentclass[aip,reprint]{revtex4-1}
\usepackage{graphicx}
\usepackage{wrapfig}
\usepackage{dcolumn}
\usepackage{bm}
\usepackage{epstopdf}
\usepackage{amsmath} 
\usepackage{amssymb} 
\usepackage[justification=raggedright]{caption}
\usepackage{textcomp}
\usepackage{tabularx} 
\usepackage{multirow}
\usepackage[ngerman, english]{babel}
\usepackage{hyperref}

\begin{document}

\preprint{APS/123-QED}

\title{Unusually high critical current of clean P-doped BaFe$_2$As$_2$ single crystalline thin film}

\author{F.\,Kurth}
\email[Electronical address:\,]{fritz.kurth@ifw-dresden.de}
\affiliation{Institute for Metallic Materials, IFW Dresden, 01171 Dresden, Germany}
\affiliation{TU Dresden, 01062 Dresden, Germany}
\author{C.\,Tarantini}
\affiliation{Applied Superconductivity Center, National High Magnetic Field Laboratory, Florida State University, 2031 East Paul Dirac Drive, Tallahassee, Florida 32310, USA}
\author{V.\,Grinenko}
\affiliation{Institute for Metallic Materials, IFW Dresden, 01171 Dresden, Germany}
\author{J.\,H\"anisch}
\affiliation{Institute for Metallic Materials, IFW Dresden, 01171 Dresden, Germany}
\affiliation{Karlsruhe Institute of Technology, Institute for Technical Physics, Hermann von Helmholtz-Platz 1, 76344 Eggenstein-Leopoldshafen, Germany}
\author{J.\,Jaroszynski}
\affiliation{Applied Superconductivity Center, National High Magnetic Field Laboratory, Florida State University, 2031 East Paul Dirac Drive, Tallahassee, Florida 32310, USA}
\author{E.\,Reich}
\affiliation{Institute for Metallic Materials, IFW Dresden, 01171 Dresden, Germany}
\author{Y.\,Mori}
\author{A.\,Sakagami}
\author{T.\,Kawaguchi}
\affiliation{Department of Crystalline Materials Science, Nagoya University, Chikusa, Nagoya 464-8603, Japan}
\author{J.\,Engelmann}
\affiliation{Institute for Metallic Materials, IFW Dresden, 01171 Dresden, Germany}
\affiliation{TU Dresden, 01062 Dresden, Germany}
\author{L.\,Schultz}
\affiliation{Institute for Metallic Materials, IFW Dresden, 01171 Dresden, Germany}
\affiliation{TU Dresden, 01062 Dresden, Germany}
\author{B.\,Holzapfel}
\affiliation{Karlsruhe Institute of Technology, Institute for Technical Physics, Hermann von Helmholtz-Platz 1, 76344 Eggenstein-Leopoldshafen, Germany}
\author{H.\,Ikuta}
\affiliation{Department of Crystalline Materials Science, Nagoya University, Chikusa, Nagoya 464-8603, Japan}
\author{R.\,H{\"u}hne}
\affiliation{Institute for Metallic Materials, IFW Dresden, 01171 Dresden, Germany}
\author{K.\,Iida}
\email[Electronical address:\,]{iida@nuap.nagoya-u.ac.jp}
\affiliation{Institute for Metallic Materials, IFW Dresden, 01171 Dresden, Germany}
\affiliation{Department of Crystalline Materials Science, Nagoya University, Chikusa, Nagoya 464-8603, Japan}

\begin{abstract}

Microstructurally clean, isovalently P-doped BaFe$_2$As$_2$ (Ba-122) single crystalline thin films have been prepared on MgO (001) substrates by molecular beam epitaxy. These films show a superconducting transition temperature ($T_\mathrm{c}$) of over 30\,K although P content is around 0.22, which is lower than the optimal one for single crystals (i.e. 0.33). The enhanced $T_\mathrm{c}$ at this doping level is attributed to the in-plane tensile strain. The strained film shows high transport self-field critical current densities ($J_\mathrm{c}$) of over 6\,MA/cm$^2$ at 4.2\,K, which are among the highest for Fe based superconductors (FeSCs). In-field $J_\mathrm{c}$ exceeds 0.1\,MA/cm$^2$ at $\mu_\mathrm{0}H=35$\,T for $H\|ab$ and $\mu_\mathrm{0}H=18$\,T for $H\|c$ at 4.2\,K, respectively, in spite of moderate upper critical fields compared to other FeSCs with similar $T_\mathrm{c}$. Structural investigations reveal no defects or misoriented grains pointing to strong pinning centers. We relate this unexpected high $J_\mathrm{c}$ to a strong enhancement of the vortex core energy at optimal $T_\mathrm{c}$, driven by in-plane strain and doping. These unusually high $J_\mathrm{c}$ make P-doped Ba-122 very favorable for high-field magnet applications.
   
\begin{description}

\item[PACS numbers]
74.70.Xa, 81.15.Fg, 74.78.-w, 74.25.Sv, 74.25.F-

\end{description}

\end{abstract}

\maketitle

Electron and hole doped BaFe$_2$As$_2$ (Ba-122) compounds show high upper critical fields with low anisotropy at low temperatures.\cite{1347-4065-51-1R-010005} These properties are favorable for high-field applications and, indeed, high-performance K-doped Ba-122 wires have been fabricated by a powder-in-tube technique.\cite{Weiss_natcom2012, gao_2014_scirep} Although, the growth of partially oriented K-doped Ba-122 thin films prepared by a two-step process has been presented\cite{leeKdopedfilms}, in-situ K-doped Ba-122 thin films are difficult to realize due to the high vapor pressure of K. It has been reported that K-doped Ba-122 thin films were prepared by molecular beam epitaxy (MBE), although the films are not stable in air.\cite{TakedaMBESrBa} On the other hand, isovalently P-doped Ba-122 thin films with a superconducting transition temperature ($T_\mathrm{c}$) of around 30\,K, which is the second highest $T_\mathrm{c}$ in the Ba-122 family, have been fabricated by both pulsed laser deposition (PLD) and MBE, where films prepared by the former method showed slightly lower $T_\mathrm{c}$.\cite{Sakagami2013, Miura2013ncomms, 1882-0786-6-9-093101} Additionally, bicrystal experiments revealed a high critical current density ($J_\mathrm{c}$) of around 1\,MA/cm$^2$ even at a grain boundary angle as large as 24\textdegree\ at 4\,K,\cite{Sakagami2013} far beyond the superconducting properties of Co-doped Ba-122.\cite{lee:212505,Iida2013189,katasenatcomm2011}

\begin{figure*}[htbp]
\centering
\includegraphics[width=2\columnwidth]{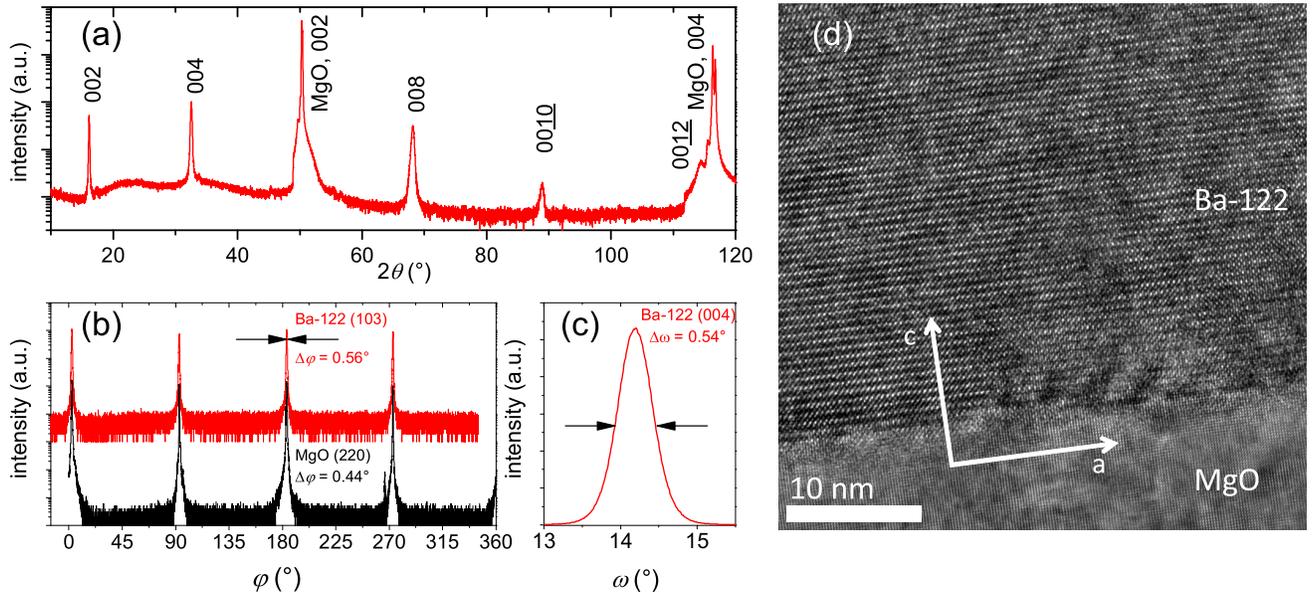}
\caption[field depending resistive measurement] {Structural data of the P-doped Ba-122 thin film. (a) $\theta-2\theta$ scan showing the $c$-axis oriented growth. (b) $\phi$-scan of the (103) reflection of the Ba-122 and the (220) reflection of the MgO substrate proving the cube-on-cube film growth. (c) Rocking curve of the (004) reflection (d) TEM image near the interface between the MgO substrate and the Ba-122 phase.}
\label{xrd}
\end{figure*}

Hence, P-doped Ba-122 may be a good candidate for Ba-122 wires and tapes applications. However, only properties at relatively low fields (up to 9\,T) have been investigated so far. In order to use this material class for applications, the knowledge of in-field transport properties in a wide range of external fields and orientations needs to be extended. Here, we present high-field (dc up to 35\,T) transport properties of a P-doped Ba-122 epitaxial thin film prepared by MBE and discuss the feasibility of applications for this material class.

\begin{figure}[htpb]
\centering
\includegraphics[width=0.98\columnwidth]{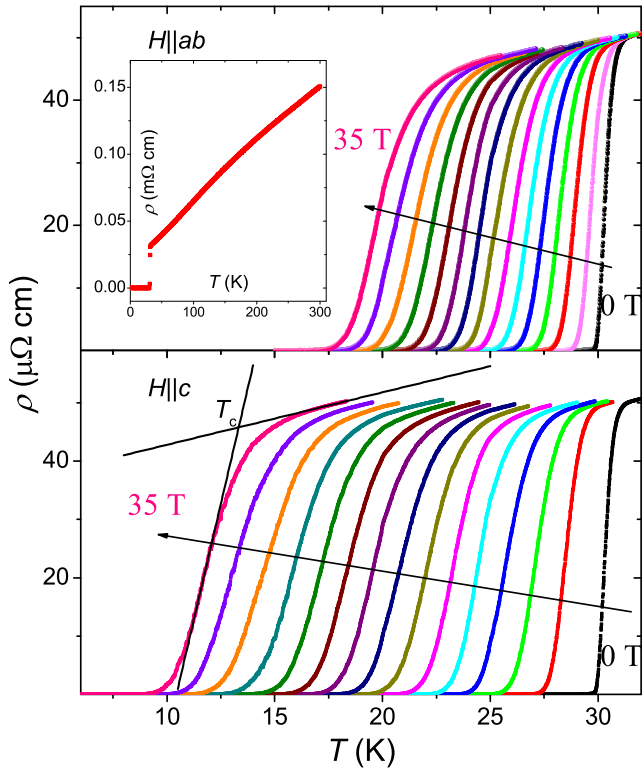}
\caption[field dependence of the resistive measurements] {Field dependence of the resistive transition for the P-doped Ba-122 thin film for $H\|ab$ (top) and $H\|c$ direction (bottom). The field steps are 2.5\,T. The inset shows the temperature dependence of the resistivity from 2\,K up to 300\,K in zero field.}
\label{rt}
\end{figure}

The P-doped Ba-122 thin film has a thickness of 107\,nm, obtained from transmission electron microscopy (TEM) and was grown on a MgO (001) substrate by MBE. The substrate was clamped to the holder with a thin indium layer in between to achieve high thermal conductivity from the heater. All elements were supplied from Knudsen cells using solid sources: Ba, Fe, As, and GaP (GaP decomposes and produces a Ga-free film). The detailed fabrication process of the thin film can be found in Ref.\citenum{0953-2048-27-6-065005}. Electron probe micro analysis was used to determine the film stoichiometry and it results to be Ba:Fe:(As+P) = 1 : 1.92 : 1.93 with a P content of P/(P+As) = 0.22, which is lower than the optimal doping level for single crystals (i.e., 0.33).\cite{PhysRevB.81.184519, 0953-8984-21-38-382203}
X-ray diffraction shows the \textit{c}-axis oriented growth of the film with high phase purity (Fig.\,\ref{xrd}a).
The (103) $\Phi$-scan (Fig.\,\ref{xrd}b) and the pole figure measurements (not shown in this letter) of the film revealed an in-plane orientation with a full width at half maximum (FWHM) value $\Delta\phi$ of 0.56\textdegree. Rocking curve measurements on the (004) reflection show a sharp peak with a FWHM value $\Delta\omega$ of 0.54\textdegree\ as can be seen in Fig.\,\ref{xrd}c. (Values are not corrected for device broadening)
The epitaxial relation was confirmed to be (001)[100]Ba-122$||$(001)[100]MgO. The overview brightfield TEM image (Fig.\,\ref{xrd}d) was taken on an FEI Tecnai T20 (LaB$_6$, 200 kV). High-resolution imaging was carried out using an FEI Titan 80-300, operating at 300 kV (field emission gun) with an image $C_\mathrm{s}$ corrector. The lamella was prepared with the in-situ lift-out method in a focused ion beam device.\cite{JEMT:JEMT20324}

In Fig.\,\ref{xrd}d no reaction layer can be seen at the interface between P-doped Ba-122 thin film and MgO substrate. Additionally, neither appreciable defects nor grain boundaries were observed in TEM investigations (lower magnifications images are not shown in this letter), indicating high crystalline quality and phase purity of the investigated films.

High field transport measurements on a small bridge of 1\,mm length and 40\,$\mu$m width, prepared by laser cutting, were conducted in the dc facility up to 35\,T at the National High Magnetic Field Laboratory (NHMFL) in Tallahassee, FL, at 4.2\,K. A criterion of 1\,$\mu$V/cm was used to define $J_\mathrm{c}$. The higher temperature measurements were performed in a 16\,T Physical Properties Measurement System. The magnetic field $H$ was applied in maximum Lorentz force configuration during all measurements (\textit{H} $\perp$ \textit{J}, \textit{J} being the current density). 
Resistivity measurements in the absence of magnetic field unveiled an onset $T_\mathrm{c}$ of 30.7\,K, which is higher than that of PLD processed films\cite{1882-0786-6-9-093101,Adachi:2012:0953-2048:105015} and comparable with the $T_\mathrm{c}$ of optimally doped single crystals.\cite{2014arXiv1402.1323P} However, the film has a lower P content than optimally doped single crystals. The lattice parameter $c = 12.78\ \mathrm{\AA}$ of our film (evaluated from XRD analysis presented in Fig.\,\ref{xrd}a, using the Nelson-Riley fit) is shorter than that of single crystals with the same P content ($c = 12.88\ \mathrm{\AA}$).\cite{PhysRevB.81.184519} This arises from the presence of tensile strain in our films. It has been reported that tensile strain in the underdoped regime for P-doped Ba-122 enhances $T_\mathrm{c}$.\cite{0953-2048-27-6-065005} The normal state resistivity shows a linear temperature dependence below 100\,K (inset of Fig.\,\ref{rt}), which is a typical behavior for optimally P-doped Ba-122.\cite{doi:10.1146/annurev-conmatphys-031113-133921} Therefore, the tensile strain shifts the superconducting transition of our film to higher values\cite{0953-2048-27-6-065005} comparable to a $T_\mathrm{c}$ of higher P concentrations without strain, a dependency which has been observed for Co-doped Ba-122 thin films as well.\cite{kurth1}

\begin{figure}[htbp]
\centering
\includegraphics[width=\columnwidth]{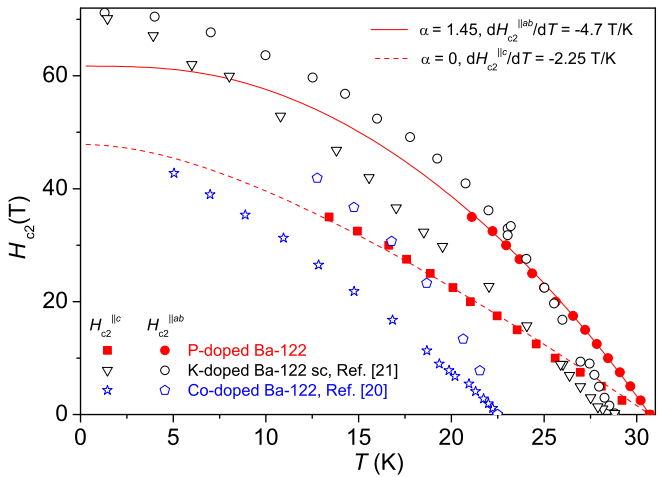}
\caption[Temperature dependence of the upper critical field]{Temperature dependence of the upper critical field of P-doped Ba-122 (red filled symbols) in $H\|c$ (squares) and $H\|ab$ (circles) direction. The data for a Co-doped Ba-122 thin film\cite{haenisch10} (blue stars and pentagon) and a K-doped Ba-122 single crystal\cite{PhysRevB.84.184522} (black triangles and circles) of comparable $T_\mathrm{c}$ are shown as well. The red lines are the single-band WHH fits of the P-doped Ba-122.}
\label{Hc2}
\end{figure}

The temperature dependence of the resistivity $\rho(T)$ was measured in magnetic fields up to 35\,T in both main crystallographic directions (Fig.\,\ref{rt}). 
The small broadening of the transition width with increasing magnetic field is similar to other Ba-122 systems, due to the small Ginzburg number.\cite{PhysRevB.85.184525} 
The temperature dependence of the upper critical field ($H_\mathrm{c2}$) values evaluated from the resistivity data shown in Fig.\,\ref{rt} is plotted in Fig.\,\ref{Hc2}. 
The $H_\mathrm{c2}$ data for both directions (i.e., $H\|c$ and $H\|ab$) can be fitted by the single-band Werthamer-Helfand-Hohenberg (WHH) model.\cite{PhysRev.147.295} The temperature dependence of $H_\mathrm{c2}$ for $H\|c$ does not show any paramagnetic limiting effects up to 35\,T in accord with literature data\cite{2014arXiv1402.1323P} (i.e., Maki parameter $\alpha=0$). However, $H_\mathrm{c2}$ for $H\|ab$ is reduced at low temperatures compared to an extrapolation from the slope at $T_\mathrm{c}$ with $\alpha$=0. In this case we obtained a value of $\alpha$\,=\,1.45. This can be attributed to Pauli-limiting behavior which is typical for FeSCs.\cite{vadim11,fuchs09} As can be seen in Fig.\,\ref{Hc2}, $H_\mathrm{c2}(T)$ for P-doped Ba-122 is relatively  small compared to other compounds having comparable $T_\mathrm{c}$.\cite{haenisch10,PhysRevB.84.184522,yuan_nature_457}  Additionally, the $H_\mathrm{c2}$ anisotropy of P-doped Ba-122 is larger than that of Co-doped and K-doped Ba-122. In general, the critical current densities are very high for P-doped Ba-122 compared to other FeSCs, as shown below.

\begin{figure*}[htpb]
\centering
\includegraphics[width=2\columnwidth]{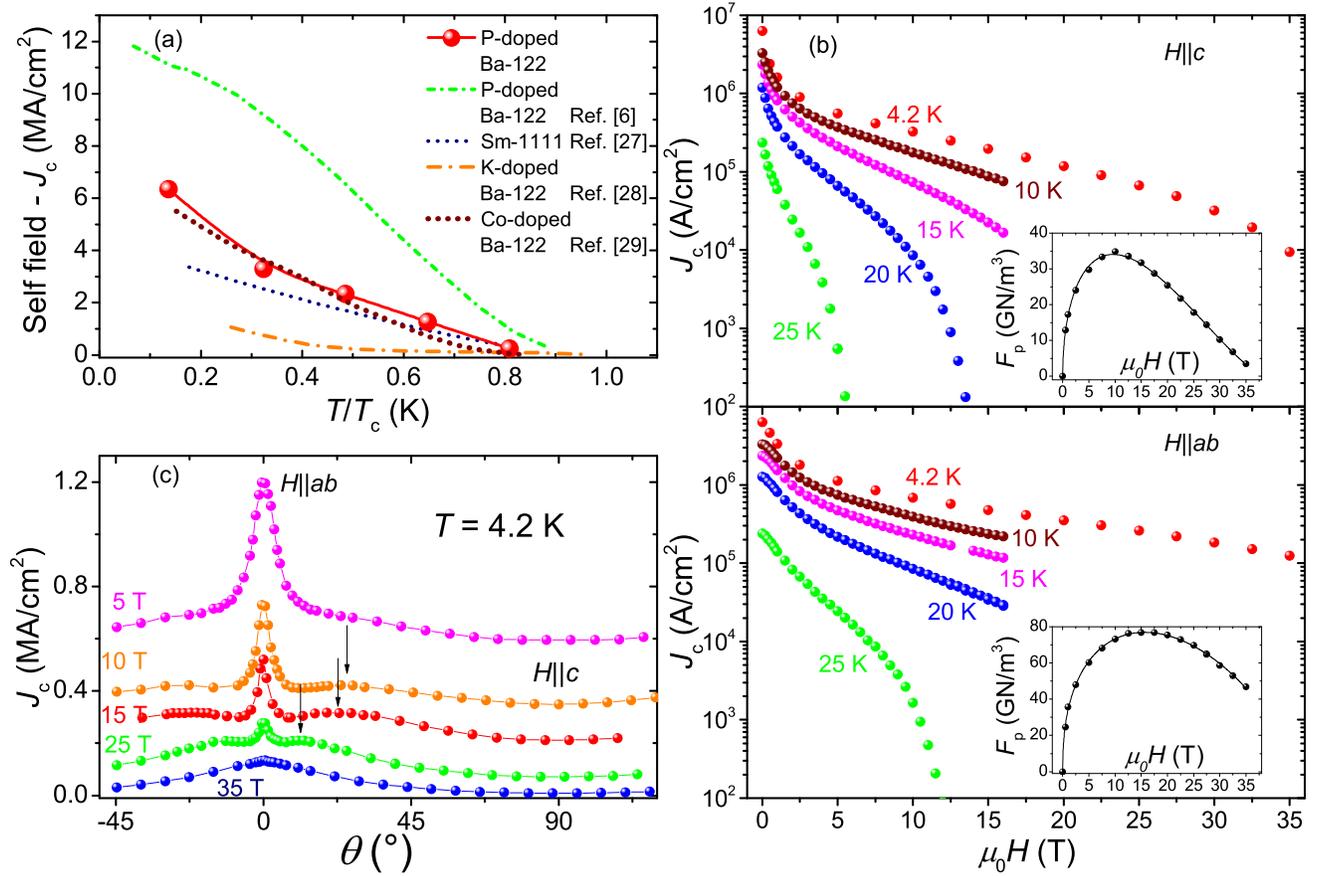}
\caption[field dependence of $J_\mathrm{c}$] {(a) Self-field $J_\mathrm{c}$ in dependence on the normalized temperature for several FeSCs.\cite{iidascirep2013,PhysRevB.80.144515,Sakagami2013,TarantiniCo-dopedCaF2} The temperature is normalized to $T_\mathrm{c,onset}$, for Sm-1111 to $T_\mathrm{c,90}$. (b) Field dependence of $J_\mathrm{c}$ of P-doped Ba-122 for both major crystallographic directions. The insets show the respective pinning force densities at 4.2\,K. (c) Angle dependence of $J_\mathrm{c}$ for different applied fields at 4.2\,K.}
\label{Jc_sf}
\end{figure*}
  
In Fig.\,\ref{Jc_sf}a the temperature dependence of the self-field $J_\mathrm{c}$ for our P-doped Ba-122 film is compared to other $J_\mathrm{c}$ results reported for FeSCs. It is seen that P-doped Ba-122 shows the highest self-field $J_\mathrm{c}$ values among other FeSCs. Note, that even SmFeAs(O,F) (Sm-1111) with a much higher \textit{T}$_\mathrm{c}$ of 54\,K shows smaller \textit{J}$_\mathrm{c}$ values.\cite{iidascirep2013,PhysRevB.80.144515,TarantiniCo-dopedCaF2} Our clean P-doped Ba-122 film presents a self-field $J_\mathrm{c}$ of about 6.3\,MA/cm$^2$ at 4.2\,K, which is almost 7\% of the depairing current density.\cite{PhysRevB.85.184525, PhysRevB.81.220501} We assume that this $J_\mathrm{c}$ could be further increased by adding additional pinning centers.\cite{Miura2013ncomms} As can be seen in Fig.\,\ref{Jc_sf}a, the film prepared by the same method but with an excess of Fe content during the growth has twice the value of $J_\mathrm{c}$.\cite{Sakagami2013}
The field dependence of $J_\mathrm{c}$ at various temperatures is summarized in Fig.\,\ref{Jc_sf}b. The $J_\mathrm{c}$ values for $H\|c$ ($J_{\mathrm{c},H\|c}$) are always lower than those for $H\|ab$, and no feature close to $H\|c$ is observed in Fig.\,\ref{Jc_sf}c: these findings indicate the absence of $c$-axis correlated pinning and that the material anisotropy dominates the general $J_\mathrm{c}$ trend. The pinning force density ($F_\mathrm{p}$, insets of Fig.\,\ref{Jc_sf}b) calculated according to $F_\mathrm{p}=H\,\times\,J_\mathrm{c}$ at 4.2\,K shows values up to 77\,GN/m$^3$ (at 15\,T) for \textit{H}$\|$\textit{ab}. The data for \textit{H}$\|$\textit{c} show a maximum of around 35\,GN/m$^3$ at around 10\,T. Compared to the results presented by Miura et al.\cite{1882-0786-6-9-093101,Miura2013ncomms} and Adachi et al.\cite{Adachi:2012:0953-2048:105015}, our film showed slightly higher values which might be due to the higher $T_\mathrm{c}$. In particular $J_\mathrm{c}$ at 35\,T $H\|c$ is as high as $J_{\mathrm{c},H\|c}$=$1\cdot 10^4$\,A/cm$^2$.
In general, $J_\mathrm{c}$ of optimally P-doped films is quite robust against applied magnetic fields.
The question arises why our microstructurally clean film exhibits such high \textit{J}$_\mathrm{c}$ values. Usually, a high density of defects is necessary to achieve high \textit{J}$_\mathrm{c}$. However, we did observe neither crystal structure defects nor impurity phases in our films (Fig.\,\ref{xrd}). Alternatively, it has been shown by Putzke et al. that the vortex core energy of the flux lines is enhanced close to the optimal doping.\cite{2014arXiv1402.1323P} Therefore, we suppose that this high vortex core energy is a key factor responsible for the unusually high \textit{J}$_\mathrm{c}$ in optimally P-doped Ba-122. In this context, comparable \textit{J}$_\mathrm{c}$ values of the P-doped Ba-122 with $T_\mathrm{c}$ around 25\,K containing BaZrO$_\mathrm{3}$ particles\cite{Miura2013ncomms} or strong pinning centers\cite{:/content/aip/journal/apl/104/18/10.1063/1.4875956} could be explained by the reduction of the vortex core energy due to non optimal \textit{T}$_\mathrm{c}$. 

Figure\,\ref{Jc_sf}c shows the angular dependence of $J_\mathrm{c}$ [$J_\mathrm{c}(\theta)$] measured at 4.2\,K and various fields up to 35\,T. $J_\mathrm{c}$ has a broad maximum positioned at $\theta=0$\textdegree\ ($H\|ab$) and no prominent $J_\mathrm{c}$ peaks at $\theta=90$\textdegree ($H\|c$). Low \textit{J}$_\mathrm{c}$(\textit{H})-anisotropy values ($\gamma_{J_\mathrm{c}}=J_\mathrm{c,H\|ab}/J_\mathrm{c,H\|c}$) of around 2 up to 15\,T at 4.2\,K are observed, increasing for higher fields. Noteworthy is the observation of a small shoulder near the $ab$-peak, marked with arrows. It shifts to lower angles when the magnetic field is increased. Such shoulders are known from cuprates and exist usually due to strong correlated defects, such as seen in double-perovskite-doped YBa$_\mathrm{2}$Cu$_\mathrm{3}$O$_\mathrm{7}$ (YBCO) thin films\cite{0953-2048-24-9-095012}, or due to uncorrelated extended defects, such as in Au-irradiated YBCO thin films\cite{:/content/aip/journal/jap/114/23/10.1063/1.4849956} or in Sm$_\mathrm{1+x}$Ba$_\mathrm{2-x}$Cu$_\mathrm{3}$O$_\mathrm{7-d}$ thin films with Sm-rich precipitates.\cite{Awaji_2007_Pc} There, the shoulders are accompanied by extended structural defects and often also by a strong $c$-axis peak in a certain magnetic field range. For our films, we did not observe any extended or correlated defects in TEM images nor any sign of off-axis peak by XRD. Therefore, the reason for these shoulders may lie in a possible variation of the P-content. For example, nanoscale regions of non-optimal P-content (not observable in TEM) of size slightly larger than the coherence lengths might act as uncorrelated strong pinning centers due to the unusually strong P-content dependence of the vortex core energy.\cite{2014arXiv1402.1323P,hashimoto_science_2012_336} These possible P-inhomogeneities in combination with the inequality of $\lambda$ and $\xi$ anisotropy as in the FeSCs can lead to such shoulders, as was shown by van der Beek et al.\cite{0953-2048-25-8-084010} The presence of the pinning centers is further evidenced by the $J_\mathrm{c}(T)$ dependence at intermediate and high temperatures (Fig.\,\ref{Jc_sf}a). Additional artificial disorder in the form of Fe impurities enhances the strong pinning resulting in a $J_\mathrm{c}(T)$ dependence consistent with the $\delta$l pinning scenario in the whole temperature range.\cite{PhysRevLett.72.1910} Thus, we believe pinning driven by differences in the vortex core energy enhances $J_\mathrm{c}$  values of the optimally P-doped Ba-122 well above all other FeSCs.

There is still room for further increase in $J_\mathrm{c}$ of the P-doped Ba-122 thin films, like doping with pinning-promoting particles the way it was done by Miura et al.\cite{Miura2013ncomms} in combination with optimal growth\cite{engelmann2013natcom,:/content/aip/journal/apl/104/18/10.1063/1.4875956}, and high crystalline quality. Additionally, a high concentration of a secondary phase like Fe can be incorporated into the superconducting matrix as artificial pinning centers for $J_\mathrm{c}$ increase without detrimental decrease in $T_\mathrm{c}$.\cite{Sakagami2013} This implies that the $J_\mathrm{c}$ anisotropy can be reduced while maintaining high $T_\mathrm{c}$ as well as $J_\mathrm{c}$. The unusually high critical currents makes the P-doped Ba-122 one of the most promising materials among FeSCs for the study of the superconducting pairing mechanisms and high field applications.

To conclude, using MBE we have fabricated P-doped Ba-122 thin film directly on MgO achieving epitaxy and phase purity with a high $T_\mathrm{c}$ of 30.7\,K. We measured the field and the angle dependence of $J_\mathrm{c}$ up to 35\,T. A very high self field $J_\mathrm{c}$ of 6.3\,MA/cm$^2$ at 4.2\,K was observed even though no structural defects were found in TEM and XRD. This observation suggests that in the optimally doped P-doped Ba-122 compound rather weak structural inhomogeneities result in strong pinning centers. This unusual pinning enhancement is explained by a sharp maximum in the vortex core energy near to the optimal doping. 

The research leading to these results has received funding from the European Union's Seventh Framework Programme (FP7/2007-2013) under agreement number 283141 (IRON-SEA) and the EU-Japan project (No.\,283204 SUPER-IRON) for support. A portion of this work was performed at the National High Magnetic Field Laboratory, which is supported by National Science Foundation Cooperative Agreement No.\,DMR-1157490, the State of Florida, and the U.S. Department of Energy. This research has also been supported by the Strategic international Collaborative Research Program (SICORP), Japan Science and Technology Agency and by a Grant-in-Aid from MEXT, Japan. J.E. acknowledges the graduate school GRK 1621.

\bibliographystyle{apsrev4-1}
\bibliography{rev}

\end{document}